\documentclass[11pt]{article}
\usepackage{amssymb,amsmath,epsfig}
\def\one{{{{\rm 1} \kern -.19em {\rm l}}}}
\def\C{{{{\rm {\mbox{\small l}}} \kern -.50em {\rm C}}}}
\def\R{{{{\rm l} \kern -.15em {\rm R}}}}
\def\N{{{{\rm l} \kern -.15em {\rm N}}}}
\def\E{{{{\rm l} \kern -.15em {\rm E}}}}
\def\P{{{{\rm l} \kern -.15em {\rm P}}}}
\def\Z{{{{\rm Z} \kern -.35em {\rm Z}}}}
\def\1{{{{\rm 1} \kern -.35em {\rm 1}}}}

\begin{document}
\begin{sloppypar}
\vspace*{0cm}
\begin{center}
{\setlength{\baselineskip}{1.0cm}{ {\Large{\bf 
DARBOUX PARTNERS OF PSEUDOSCALAR \\ DIRAC POTENTIALS ASSOCIATED WITH \\ EXCEPTIONAL 
ORTHOGONAL POLYNOMIALS
\\}} }}
\vspace*{1.0cm}
{\large{\sc{Axel Schulze-Halberg}}$^\ddagger$ and {
\\[1ex]
\sc{Barnana Roy}}$^\dagger$}
\end{center}
\noindent \\
$\ddagger$ Department of Mathematics and Actuarial Science and Department of Physics, Indiana University Northwest, 3400 Broadway,
Gary IN 46408, USA, e-mail: axgeschu@iun.edu, xbataxel@gmail.com
 \\ \\
$\dagger$ Physics \& Applied Mathematics Unit, Indian Statistical Institute, Kolkata 700108, India,
e-mail: barnana@isical.ac.in

\vspace*{.5cm}
\begin{abstract}
\noindent
We introduce a method for constructing Darboux (or supersymmetric) pairs of pseudoscalar and scalar 
Dirac potentials that are associated with exceptional orthogonal polynomials. Properties of 
the transformed potentials and regularity conditions are discussed. As an application, 
we consider a pseudoscalar Dirac potential related to the Schr\"odinger model for the 
rationally extended radial oscillator. The pseudoscalar partner potentials are constructed under first-
and second-order Darboux transformations.

\end{abstract}

\section{Introduction}
\noindent
In recent years, a considerable amount of research has been devoted to studying exceptional $X_l$
orthogonal polynomials that lead to rational extensions of quantum mechanical potentials.
Exceptional orthogonal polynomials (EOPs) of codimension one were first introduced in \cite{Ullate1}
as the polynomial eigenfunctions of a Sturm-Liouville problem. In contrast to the families of
classical orthogonal polynomials, which start at degree $0$, the exceptional ones start at
degree $m (m\geq 1)$ but still form complete sets with respect to some positive definite
measure. The relationship between exceptional orthogonal polynomials, the Darboux
 transformation and shape invariant potentials was shown in \cite{Quesne1} followed
by \cite{Quesne2} . Higher codimensional families were obtained in \cite {Odake1}.
Multi-indexed families associated to Darboux-Crum or iterated Darboux transformations
were studied in \cite{Odake3}. Along with the study of the mathematical properties of
these exceptional polynomial families \cite{Ullate2}, the latter has found applications in a
number of physical problems, e.g. quantum superintegrability \cite{Quesne3}, Dirac
operators minimally or non minimally
coupled to external fields and Fokker Planck equations \cite{Ho1}, entropy measures in
quantum information theory \cite{Dutta}, solutions of Schr$\ddot{o}$dinger equation with
some conditionally exactly solvable potentials \cite{Roy}, solutions for position dependent
mass systems \cite{Midya}, discrete quantum mechanics \cite{Sasaki}, quantum Hamilton-Jacobi
formalism \cite{Sree}, $\cal{N}$-fold Supersymmetry and its dynamical breaking in the context of
position dependent mass \cite{tanaka}.

In this note our objective is to show, in principle, how to obtain via Darboux transformation,
exactly solvable scalar and pseudoscalar potentials and their supersymmetric partners for
(1+1)-dimensional Dirac equation associated to exceptional orthogonal polynomials.
It should be mentioned here that a Darboux transformation for Dirac equation has been
constructed for both one and two dimensional stationary case \cite{Samsonov1} and the
 fully time-dependent equation in (1+1) dimensions \cite{Samsonov2}. The motivation of the
present work stems from the fact that with the experimental realization of graphene \cite{Novo} and
of the massless Dirac nature of its electron low-energy spectrum, a renovated interest in obtaining
analytical solutions for the energy eigenvalues and eigenfunctions for electrons and holes of the
 massless or mass dependent on position Dirac equation in (2+1) as well as in (1+1) dimensions
has started to emerge \cite{Imsc}.
%Considering the recent surge of interest in exceptional orthogonal
%polynomials and scarcity of literature [8] so far Dirac equation is concerned, this type of study
%certainly deserves attention.

The remainder of this article is organized as follows. In section 2 we demonstrate how
one-parameter classes of pseudoscalar and scalar Dirac potentials can be generated by
starting out from a solvable Schr\"odinger equation. Section 3 is devoted to the Darboux
transformation for the Dirac equation, where we will mainly be summarizing results from
\cite{Samsonov1}. In section 4 we apply the Darboux formalism to a particular class of
pseudoscalar Dirac potentials that arise from a Schr\"odinger equation for a rationally extended
oscillator potential, while section 5 contains a final discussion concerning regularity of the 
transformed potentials and related issues.

\section{Construction of Dirac potentials}
We start out by considering the conventional stationary Schr\"odinger equation in atomic units
\begin{eqnarray}
\psi''+(\epsilon-V_0)~\psi &=& 0, \label{sse}
\end{eqnarray}
where $\epsilon$ is the real-valued energy and the continuous function $V_0$ stands for the potential.
Let us assume that $\psi$ is a solution of the Schr\"odinger equation (\ref{sse}).
For our subsequent considerations it is essential to rewrite the potential $V_0$ in the following form
\begin{eqnarray}
V_0 &=& q_0^2 + q_0', \label{v0q0}
\end{eqnarray}
introducing a differentiable function $q_0$. For a given potential $V_0$, the general solution to the
Riccati equation (\ref{v0q0}) determines a one-parameter family of functions $q_0$ that we will use to
construct a pseudoscalar or scalar Dirac potential. Before we do so, let us elaborate on how to find the
general solution of (\ref{v0q0}). It is well-known \cite{kamke} that the latter solution can be constructed once a
particular solution is known. We can come by such a particular solution after linearizing (\ref{v0q0}) via
$q_0=\hat{q}_0'/\hat{q}_0$, leading to the result
\begin{eqnarray}
\hat{q}_0''-V_0~\hat{q}_0 &=& 0. \label{v0q0lin}
\end{eqnarray}
Comparison of this equation and (\ref{sse}) shows that $\hat{q}_0=\psi_{|\epsilon=0}$ is a solution of
(\ref{v0q0lin}), such that
\begin{eqnarray}
q_{0,p} &=& \frac{d}{dx}~ \log\left(\hat{q}_0\right) ~=~ \frac{d}{dx}~ \log\left(\psi_{|\epsilon=0}\right), \label{q0p}
\end{eqnarray}
is a particular solution to our Riccati equation (\ref{v0q0}). Once we have found (\ref{q0p}), we can
determine the general solution $q_0$ of (\ref{v0q0}) through the formula \cite{kamke}
\begin{eqnarray}
q_0 &=& q_{0,p} + \frac{\exp\left(-2~\int\limits^x q_{0,p}~dt \right)}{c+\int\limits^x
\exp\left(-2~\int\limits^t q_{0,p}~ds \right)~dt}, \label{q1pp}
\end{eqnarray}
where $c$ is an arbitrary constant. Expression (\ref{q1pp}) simplifies if we insert the
form of $q_{0,p}$ given in (\ref{q0p}):
\begin{eqnarray}
q_0 &=&  \left[ \frac{d}{dx}~ \log\left(\hat{q}_0 \right) \right] + \left(
c~\hat{q}_0^2 + \hat{q}_0^2 \int\limits^x \frac{1}{\hat{q}_0^2}~dt
\right)^{-1}, \label{q0final}
\end{eqnarray}
recall that $\hat{q}_0=\psi_{|\epsilon=0}$. As mentioned before, (\ref{q0final}) is a one-parameter family
that will now be implemented as the parametrizing function of a Dirac potential. To this end, let us
distinguish the following two cases.

\paragraph{Pseudoscalar case.} First, we use (\ref{q0final}) and the
solution $\psi$ to the Schr\"odinger equation (\ref{sse}) for defining two functions $\Psi_1$ and
$\Psi_2$ as follows
\begin{eqnarray}
\Psi_1 &=& \psi \label{p1} \\[1ex]
\Psi_2 &=& \frac{1}{E+m} \left(q_0~\Psi_1-\Psi_1' \right), \label{p2}
\end{eqnarray}
where $m$ is a positive constant and $|E|=\sqrt{\epsilon+m^2}$. The function $\Psi=(\Psi_1,\Psi_2)^T$
is a solution of the Dirac equation
\begin{eqnarray}
i~\sigma_2~\Psi' + (U_0 - E)~\Psi &=& 0. \label{dirac}
\end{eqnarray}
Here, $\sigma_2$ denotes the usual Pauli Matrix and the potential is given by
\begin{eqnarray}
U_0 &=& m~\sigma_3+q_0~\sigma_1, \label{v0}
\end{eqnarray}
for Pauli matrices $\sigma_1$ and $\sigma_3$. Expression (\ref{v0}) represents a pseudoscalar
potential for the Dirac equation (\ref{dirac}). In conclusion, we have converted a solution $\psi$ of
our initial Schr\"odinger equation to a solution of the Dirac equation (\ref{dirac}) for the
pseudoscalar potential (\ref{v0}).

\paragraph{Scalar case.} We start out by defining a function $S_0$ as follows:
\begin{eqnarray}
S_0 &=& q_0+m, \label{s0}
\end{eqnarray}
where $q_0$ is given in (\ref{q0final}). Next, we introduce two functions
$\Psi_1$ and $\Psi_2$ by
\begin{eqnarray}
\Psi_1 &=& \psi \label{p1s} \\[1ex]
\Psi_2 &=& \frac{1}{E} \left[(m+S_0)~\Psi_1-\Psi_1' \right], \label{p2s}
\end{eqnarray}
for a positive constant $m$ and $|E|=\sqrt{\epsilon}$. The function $\Psi=(\Psi_1,\Psi_2)^T$ is a solution
of the Dirac equation
\begin{eqnarray}
i~\sigma_2~\Psi' + (U_0 - E)~\Psi &=& 0, \label{diracs}
\end{eqnarray}
where the potential has scalar form
\begin{eqnarray}
U_0  &=& (m+S_0)~\sigma_1. \label{vsca}
\end{eqnarray}
Thus, relations (\ref{p1s}), (\ref{p2s}) establish a link between the solutions of the Schr\"odinger
equation (\ref{sse}) and its Dirac counterpart (\ref{diracs}) for the scalar potential (\ref{vsca}). Recall
that the function $S_0$ can be given in explicit form, once we combine (\ref{s0}) and (\ref{q0final}).

\section{Darboux transformation for the Dirac equation}
In this section we will explain how the Darboux transformation can be applied to a Dirac equation for
potentials of pseudoscalar or scalar form. We will summarize particular results from \cite{Samsonov1}, the
reader may refer to the latter reference for a more detailed discussion of the topic. We consider a Dirac
equation in the form
\begin{eqnarray}
i~\sigma_2~\Psi' + (V_0 - E)~\Psi &=& 0, \label{diracd}
\end{eqnarray}
where $V_0$ takes either pseudoscalar (\ref{v0}) or scalar form (\ref{vsca}). Note that in contrast to the
previous section, the parametrizing
functions $q_0$ and $S_0$ are allowed to remain undetermined for now. If we assume that
$\Psi=(\Psi_1,\Psi_2)^T$ is a solution to our Dirac equation (\ref{diracd}), then its first component
$\Psi_1=\psi$ solves a stationary Schr\"odinger equation of the form
\begin{eqnarray}
\psi''+(\epsilon-q_0^2-q_0')~\psi &=& 0. \label{ssed}
\end{eqnarray}
If the potential $V_0$ in our initial Dirac equation is pseudoscalar, then in (\ref{ssed}) we have
$\epsilon=E^2-m^2$. In case of a scalar potential $V_0$, the function $q_0$ is defined through
(\ref{s0}), and $\epsilon=E^2$. We will now apply the Darboux transformation to equation (\ref{ssed}). To this
end, let $u_1,...,u_{N}$ be $N$ auxiliary solutions of (\ref{ssed}) for the respective, pairwise different
energy values $\lambda_1,...,\lambda_N$. Then,
\begin{eqnarray}
\phi &=& \frac{W_{u_1,...,u_N,\psi}}{W_{u_1,...,u_N}}, \label{phidarb}
\end{eqnarray}
where $W$ stands for the Wronskian of the functions denoted in its index, is a solution of the Schr\"odinger
equation
\begin{eqnarray}
\phi''+\left[\epsilon-q_0^2-q_0'+2~\frac{d}{dx} \log\left(W_{u_1,...,u_N} \right) \right] \phi &=& 0.
\label{ssedt}
\end{eqnarray}
We must now find an interrelation between (\ref{ssedt}) and a Dirac equation of the form (\ref{diracd}).
To this end, we impose the condition
\begin{eqnarray}
q_0^2+q_0'-2~\frac{d}{dx} \log\left(W_{u_1,...,u_N} \right)  &=& q_1^2+q_1', \label{q0wq1}
\end{eqnarray}
which represents a Riccati equation for the new function $q_1$. This function can be found in the same way as
it was done for its counterpart $q_0$ in (\ref{v0q0}). Adopting the notation from (\ref{q0p}),
we find that a particular solution $q_{1,p}$ to (\ref{q0wq1}) is given by
\begin{eqnarray}
q_{1,p} &=&  \frac{d}{dx} ~\log\left( \hat{q}_1\right) ~=~
 \frac{d}{dx} ~\log\left(\phi_{\mid \epsilon = 0} \right). \label{q1p}
\end{eqnarray}
Recall that $\phi$ is known from (\ref{phidarb}). The general solution of (\ref{q0wq1}) can be
constructed by means of the formula
\begin{eqnarray}
q_1 &=&  \left[ \frac{d}{dx}~ \log\left(\hat{q}_1 \right) \right] + \left(
c~\hat{q}_1^2 + \hat{q}_1^2 \int\limits^x \frac{1}{\hat{q}_1^2}~dt
\right)^{-1}. \label{q1final}
\end{eqnarray}
Now that $q_1$ has been determined, we are ready to state the transformed Dirac equation as well as
its solutions. Starting out with the equation, it reads
\begin{eqnarray}
i~\sigma_2~\Phi' + (U_1 - E)~\Phi &=& 0, \label{diracdt1}
\end{eqnarray}
where the potential $U_1$ has either pseudoscalar or scalar form. In the first case, the potential
is given by
\begin{eqnarray}
U_1 &=& m~\sigma_3+q_1~\sigma_1, \label{u1}
\end{eqnarray}
the function $q_1$ being stated in (\ref{q1final}),
while the corresponding solution $\Phi=(\Phi_1,\Phi_2)^T$ of (\ref{diracdt1})
takes the same shape as in (\ref{p1}), (\ref{p2}), that is,
\begin{eqnarray}
\Phi_1 &=& \frac{W_{u_1,...,u_N,\Psi_1}}{W_{u_1,...,u_N}} \nonumber \\[1ex]
\Phi_2 &=& \frac{1}{E+m} \left(q_1~\Phi_1-\Phi_1' \right), \nonumber
\end{eqnarray}
recall that $\phi$ is defined in (\ref{phidarb}). If the potential $U_1$ takes scalar form, it reads
\begin{eqnarray}
U_1  &=& (m+S_1)~\sigma_1, \nonumber
\end{eqnarray}
introducing a new function $S_1=q_1+m$. The corresponding solution $\Phi=(\Phi_1,\Phi_2)^T$
of (\ref{diracdt1}) is obtained from (\ref{p1s}), (\ref{p2s}) as
\begin{eqnarray}
\Phi_1 &=& \frac{W_{u_1,...,u_N,\Psi_1}}{W_{u_1,...,u_N}} \nonumber \\[1ex]
\Phi_2 &=& \frac{1}{E} \left[(m+S_1)~\Phi_1-\Phi_1' \right]. \nonumber
\end{eqnarray}
In summary, we have established a connection between the initial Dirac equation (\ref{diracd}) and
its transformed counterpart (\ref{diracdt1}) by means of the Darboux transformation (\ref{phidarb}).

\section{Application: the extended radial oscillator model}
We will now combine the results from sections 2, 3 and apply them to a Dirac model that can be
generated by means of the Schr\"odinger equation for the rationally extended radial oscillator potential.
The main reasons for choosing this particular model are its simplicity and the fact that the
transformed potentials turn out to be non-singular. We will first discuss the case of a
pseudoscalar Dirac potential under first- and second-order Darboux transformations.

\begin{figure}[h]
\begin{center}
\epsfig{file=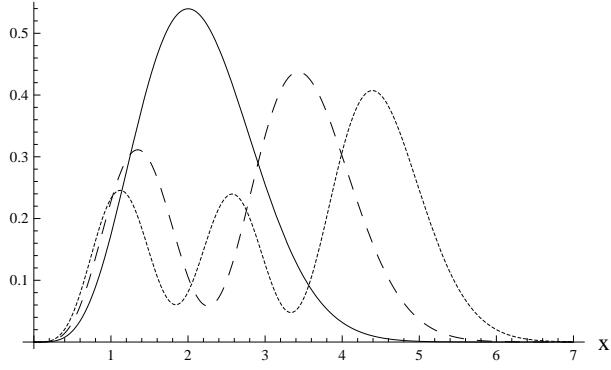,width=8cm}
\caption{Normalized probability densities $|\Psi_1|^2+|\Psi_2|^2$ for the solutions (\ref{p1rad}), (\ref{p2rad}) with
the settings $m=\omega=l=1$, $n=0$ (solid curve), $n=1$ (dashed curve), and $n=2$ (dotted curve).}
\label{prob3}
\end{center}
\end{figure}

\begin{figure}[h]
\begin{center}
\epsfig{file=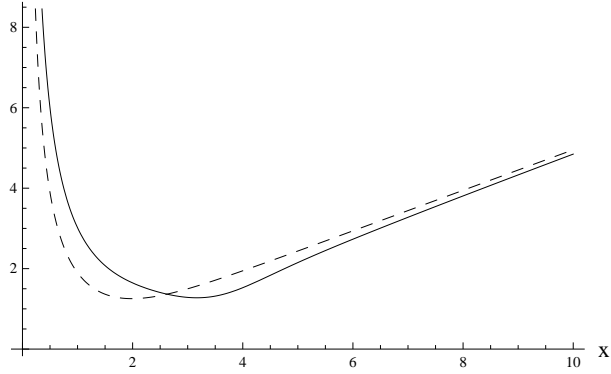,width=8cm}
\caption{The initial and transformed parametrizing functions $q_0$ (dashed curve) and $q_1$ (solid curve),
as defined in (\ref{q0rad}) and (\ref{q1finalrad}).}
\label{q0q1}
\end{center}
\end{figure}

\begin{figure}[h]
\begin{center}
\epsfig{file=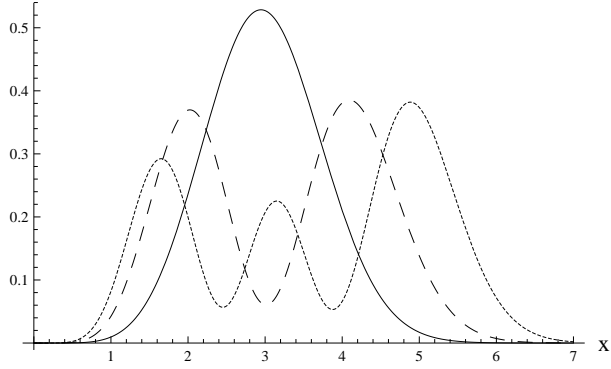,width=8cm}
\caption{Normalized probability densities $|\Phi_1|^2+|\Phi_2|^2$ for the solutions (\ref{phi1fin}), (\ref{phi2fin}) with
the settings $m=\omega=l=1$, $n=0$ (solid curve), $n=1$ (dashed curve), and $n=2$ (dotted curve).}
\label{prob3transf}
\end{center}
\end{figure}

\begin{figure}[h]
\begin{center}
\epsfig{file=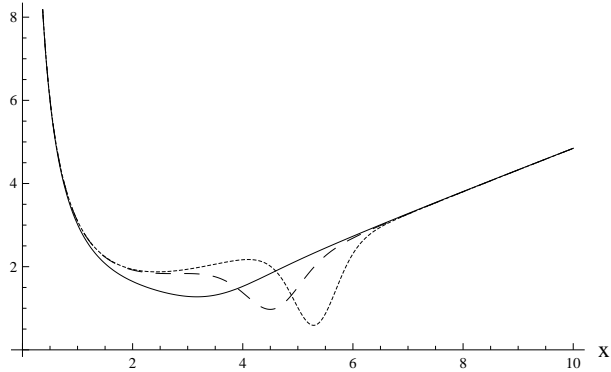,width=8cm}
\caption{The parametrizing function $q_1$ for the parameters settings $\omega=l=1$. Furthermore, we chose
the number $n$ in (\ref{eps}) as $n=-1/2$ (solid curve), $n=-1/50$ (dashed curve), and $n=-10^{-4}$
(dotted curve).}
\label{q1ene}
\end{center}
\end{figure}

\begin{figure}[h]
\begin{center}
\epsfig{file=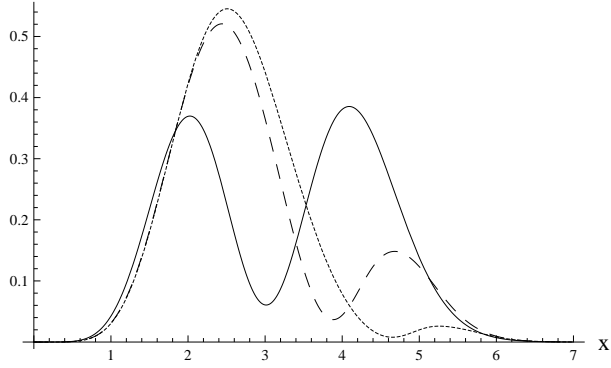,width=8cm}
\caption{Normalized probability densities $|\Phi_1|^2+|\Phi_2|^2$ for
the settings $m=\omega=l=n=1$. The energy $\lambda$ of the auxiliary solution is taken from (\ref{eps}) for
$n=-1/2$ (solid curve), $n=-1/50$ (dashed curve), and $n=-10^{-4}$ (dotted curve).}
\label{phin1ene}
\end{center}
\end{figure}

\begin{figure}[h]
\begin{center}
\epsfig{file=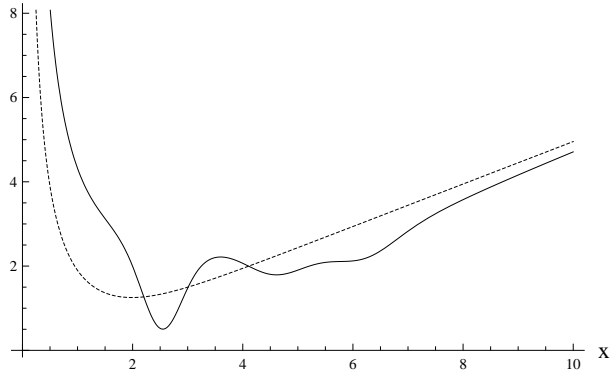,width=8cm}
\caption{The parametrizing function $q_1$ (solid curve) obtained from a second-order 
Darboux transformation, compared to its initial counterpart $q_0$ (dotted curve). We used 
the parameters settings $\omega=l=1$. The numbers $n=n_1$ and $n=n_2$ in (\ref{eps}) were chosen as 
$n_1=3/2$ and $n_2=5/4$.}
\label{q1_2nd}
\end{center}
\end{figure}

\paragraph{Pseudoscalar case.} Our starting point is the Schr\"odinger equation for the rationally extended
radial oscillator potential. The corresponding boundary-value problem has the following form
\begin{eqnarray}
\psi''(x)+\left[\epsilon-\frac{1}{4}~\omega^2~x^2-\frac{l~(l+1)}{x^2}-\frac{4~\omega}{\omega~x^2+2~l+1}+
\frac{8~\omega~(2~l+1)}{(\omega~x^2+2~l+1)^2}
\right] \psi(x) &=& 0,~ x>0 \nonumber \\
\label{bvpr1} \\[1ex]
\psi\left(0\right) ~~=~~\lim\limits_{x \rightarrow \infty} \psi(x) &=& 0, \label{bvpr2}
\end{eqnarray}
where $\omega>0$, $l$ is a nonnegative integer, and the real-valued constant $\epsilon$ represents the spectral
parameter. The problem (\ref{bvpr1}), (\ref{bvpr2}) has an infinite number of solutions $(\psi_n)$ and a
corresponding discrete spectrum $(\epsilon_n)$, where $n$ is a positive integer or zero \cite{Quesne1}
\begin{eqnarray}
\epsilon_n &=& \omega \left(2~n+l+\frac{3}{2} \right) \label{eps} \\
\psi_n(x) &=& \frac{x^{l+1}}{\omega~x^2+2~l+1}~\exp\left(-\frac{1}{4}~\omega~x^2 \right)~
{\cal L}_{n+1}^{l+\frac{1}{2}}\left(\frac{1}{2}~\omega~x^2 \right). \label{solosc}
\end{eqnarray}
Here, the symbol ${\cal L}$ stands for an exceptional Laguerre polynomial of $X_1$-type, defined as follows
\cite{Quesne1}
\begin{eqnarray}
{\cal L}_n^k(x) &=& -(x+k+1)~L_{n-1}^k(x)+L_{n-2}^k(x), \label{lx1}
\end{eqnarray}
where $L$ denotes a conventional associated Laguerre polynomial \cite{abram}. Let us now construct
a pseudoscalar potential for the Dirac equation from the boundary-value problem(\ref{bvpr1}), (\ref{bvpr2}).
To this end, we must set up and solve the Riccati equation (\ref{v0q0}), which reads in the present case
\begin{eqnarray}
\frac{1}{4}~\omega^2~x^2+\frac{l~(l+1)}{x^2}+\frac{4~\omega}{\omega~x^2+2~l+1}-
\frac{8~\omega~(2~l+1)}{(\omega~x^2+2~l+1)^2}
 &=& q_0^2 + q_0', \label{ricrad}
\end{eqnarray}
observe that we extracted the Schr\"odinger potential $V_0$ from equation (\ref{bvpr1}). Since the general
solution of equation (\ref{ricrad}) for $q_0$ involves very long expressions, we will restrict ourselves to
the particular solution $q_{0,p}$. According to (\ref{q0p}), we can construct $q_{0,p}$ by evaluating
the solution (\ref{solosc}) at the energy $\epsilon=0$. Since the latter solution does not contain
$\epsilon$ itself, we must find the numerical value for $n$ that corresponds to vanishing $\epsilon$.
Inspection of (\ref{eps}) gives this value as
\begin{eqnarray}
n &=& -\frac{l}{2}-\frac{3}{4}. \label{enull}
\end{eqnarray}
We therefore obtain the following particular solution $q_0=q_{0,p}$ of our Riccati equation (\ref{ricrad}):
\begin{eqnarray}
q_0 &=& \frac{d}{dx}~\log\left[ \frac{x^{l+1}}{\omega~x^2+2~l+1}~\exp\left(-\frac{1}{4}~\omega~x^2 \right)~
{\cal L}_{-\frac{l}{2}+\frac{1}{4}}^{l+\frac{1}{2}}\left(\frac{1}{2}~\omega~x^2 \right) \right] \nonumber \\[1ex]
&=& -\frac{\omega}{2}~x+\frac{l-1}{x}+\frac{4~l+2}{\omega~x^3+2~l~x+x}+
\omega~x~\frac{d}{du}~ \log\left[
{\cal L}_{-\frac{l}{2}+\frac{1}{4}}^{l+\frac{1}{2}}\left(u \right)  \right]_{\mid u = \omega x^2 /2}, \label{q0rad}
\end{eqnarray}
We observe that the function ${\cal{L}}$ in this expression is not a polynomial anymore, because its
lower index is not an integer.
\begin{figure}[h]
\begin{center}
\epsfig{file=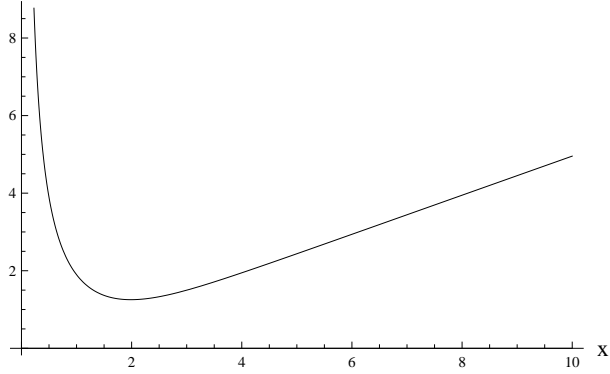,width=8cm}
\caption{The parametrizing function $q_0$, as given in (\ref{q0rad}).}
\label{q0fig}
\end{center}
\end{figure}
However, the definition (\ref{lx1}) still holds, but the associated Laguerre
polynomials turn into associated Laguerre functions \cite{abram}. Figure \ref{q0fig} shows a plot of the
function $q_0$, which parametrizes a pseudoscalar Dirac potential $U_0$ of the form (\ref{v0}):
\begin{eqnarray}
U_0 &=& m~\sigma_3+
\left\{
\frac{d}{dx}~\log\left[ \frac{x^{l+1}}{\omega~x^2+2~l+1}~\exp\left(-\frac{1}{4}~\omega~x^2 \right)~
{\cal L}_{-\frac{l}{2}+\frac{1}{4}}^{l+\frac{1}{2}}\left(\frac{1}{2}~\omega~x^2 \right) \right]
\right\}
\sigma_1. \nonumber
\end{eqnarray}
The corresponding
boundary-value problem for the Dirac equation (\ref{dirac}) reads
\begin{eqnarray}
i~\sigma_2~\Psi' + (U_0 - E)~\Psi &=& 0,~~~x>0 \label{bvpd1} \\[1ex]
\Psi(0) ~=~ \lim\limits_{x \rightarrow \infty} \Psi(x) &=& (0,0)^T. \label{bvpd2}
\end{eqnarray}
The solution $\Psi=(\Psi_1,\Psi_2)^T$ of this problem can be obtained by combining the general formulas
(\ref{p1}), (\ref{p2}) with the above solution (\ref{solosc}):
\begin{eqnarray}
\Psi_1 &=& \frac{x^{l+1}}{\omega~x^2+2~l+1}~\exp\left(-\frac{1}{4}~\omega~x^2 \right)~
{\cal L}_{n+1}^{l+\frac{1}{2}}\left(\frac{1}{2}~\omega~x^2 \right) \label{p1rad} \\[1ex]
\Psi_2 &=& \frac{1}{m+\sqrt{m^2+\omega~\left(2~n+l+\frac{3}{2}\right)}}
~\left(q_0~\Psi_1-\Psi_1' \right), \label{p2rad}
\end{eqnarray}
where we omit to substitute the explicit expressions (\ref{solosc}) and (\ref{q0rad}) for the solution and the
parametrizing function $q_0$, respectively. Before we continue, let us point out that the function
$\Psi_2$, as given in (\ref{p2rad}), in fact vanishes at positive infinity, because both the product $q_0\Psi_1$
and the derivative $\Psi_1'$ become zero there. This is illustrated in figure \ref{prob3}, where 
normalized probability densities asociated with the functions (\ref{p1rad}), (\ref{p2rad}) are displayed.
Finally, the discrete spectral values $E_n$, $n$ a nonnegative
integer, that are associated with the solutions given in (\ref{p1rad}), (\ref{p2rad}), are found from
(\ref{eps}) as
\begin{eqnarray}
|E_n| &=& \sqrt{m^2+\omega~\left(2~n+l+\frac{3}{2}\right)}. \label{isoene}
\end{eqnarray}
Now that the first of the two Dirac partners has been
determined, the remaining task is to find its Darboux partner. To this end, let us return to the initial
Schr\"odinger equation and apply the Darboux transformation to it. As a first step, we must find an auxiliary
function $u_1$ that solves our initial equation (\ref{bvpr1}) at an energy $\lambda \neq \epsilon$. We
set $n=-1/2$ in (\ref{eps}), giving
\begin{eqnarray}
\lambda &=& \omega \left(l+\frac{1}{2} \right). \label{lambda}
\end{eqnarray}
We observe from (\ref{eps}) that this energy $\lambda$ lies below the ground state energy $\epsilon_0$ of the problem
(\ref{bvpr1}), (\ref{bvpr2}). We are therefore guaranteed that the latter spectral problem and its Darboux
partner will admit the same discrete spectrum, that is, no spectral value will be added or deleted as a result of
the Darboux transformation. Next, we choose the auxiliary solution of (\ref{bvpr1}) at energy (\ref{lambda})
from the set (\ref{solosc}) by incorporating $n=-1/2$:
\begin{eqnarray}
u_1 &=& \frac{x^{l+1}}{\omega~x^2+2~l+1}~\exp\left(-\frac{1}{4}~\omega~x^2 \right)~
{\cal L}_{\frac{1}{2}}^{l+\frac{1}{2}}\left(\frac{1}{2}~\omega~x^2 \right). \label{u1rad}
\end{eqnarray}
We can now perform the first-order Darboux transformation after substituting (\ref{u1rad}) and
(\ref{solosc}) into formula (\ref{phidarb}) for $N=1$. The Darboux transformation delivers the
following function $\phi$
\begin{eqnarray}
\phi &=& -\frac{u_1'}{u_1}~\psi_n+\psi_n' \label{firsto} \\[1ex]
&=& -\frac{x^{l+2} ~\exp\left(-\frac{1}{4}~\omega~x^2 \right) \omega
~{\cal L}_{n+1}^{l+\frac{1}{2}}\left(\frac{1}{2}~\omega~x^2 \right)
}{(\omega~x^2+2~l+1)~{\cal L}_{\frac{1}{2}}^{l+\frac{1}{2}}\left(\frac{1}{2}~\omega~x^2 \right)}
\left[\frac{d}{du}~{\cal L}_{\frac{1}{2}}^{l+\frac{1}{2}}\left(u \right)\right]_{\mid u = \omega x^2 /2}
+ \nonumber \\[1ex]
&+& \frac{x^{l+2} ~\exp\left(-\frac{1}{4}~\omega~x^2 \right) \omega}{\omega~x^2+2~l+1}
\left[\frac{d}{du}~{\cal L}_{n+1}^{l+\frac{1}{2}}\left(u \right)\right]_{\mid u = \omega x^2 /2}. \label{phi1}
\end{eqnarray}
Before we relate this function to the transformed Dirac equation, let us construct the corresponding
pseudoscalar potential. To this end, we need to determine the parametrizing function $q_1$
according to (\ref{q1p}) and (\ref{q1final}). Due to the length of the expressions involved, we will
restrict ourselves to the simplest case, choosing $q_1=q_{1,p}$ as in (\ref{q1p}). The argument
$\phi_{\mid \epsilon=0}$ of the logarithm in the latter reference is obtained from (\ref{phi1}) by
setting $n$ as in (\ref{enull}):
\begin{eqnarray}
\phi_{\mid \epsilon=0} &=& \phi_{\left| n=-\frac{l}{2}-\frac{3}{4} \right.}. \label{phienull}
\end{eqnarray}
Taking into account (\ref{q1p}), we have the following expression for the parametrizing function $q_1=q_{1,p}$
that will enter in the transformed Dirac equation:
\begin{eqnarray}
q_1 &=& \frac{d}{dx}~\log\left\{
 -\frac{x^{l+2} ~\exp\left(-\frac{1}{4}~\omega~x^2 \right) \omega
~{\cal L}_{-\frac{l}{2}+\frac{1}{4}}^{l+\frac{1}{2}}\left(\frac{1}{2}~\omega~x^2 \right)
}{(\omega~x^2+2~l+1)~{\cal L}_{\frac{1}{2}}^{l+\frac{1}{2}}\left(\frac{1}{2}~\omega~x^2 \right)}
\left[\frac{d}{du}~{\cal L}_{\frac{1}{2}}^{l+\frac{1}{2}}\left(u \right)\right]_{\mid u = \omega x^2 /2}
+ \right. \nonumber \\[1ex]
&+& \left. \frac{x^{l+2} ~\exp\left(-\frac{1}{4}~\omega~x^2 \right) \omega}{\omega~x^2+2~l+1}
\left[\frac{d}{du}~{\cal L}_{-\frac{l}{2}+\frac{1}{4}}^{l+\frac{1}{2}}\left(u \right)\right]_{\mid u = \omega x^2 /2}
\right\}. \label{q1finalrad}
\end{eqnarray}
Figure (\ref{q0q1}) shows a plot of this function $q_1$, compared to its counterpart $q_0$.
The corresponding pseudoscalar potential $U_1$ is then given by (\ref{u1}), where the function
(\ref{q1finalrad}) is to be inserted for $q_1$. Next, let us set up the transformed boundary-value
problem for the Dirac equation, which takes almost the same form as (\ref{bvpd1}),
(\ref{bvpd2})
\begin{eqnarray}
i~\sigma_2~\Phi' + (U_1 - E)~\Phi &=& 0,~~~x>0 \label{bvpd3} \\[1ex]
\Phi(0) ~=~ \lim\limits_{x \rightarrow \infty} \Phi(x) &=& (0,0)^T. \label{bvpd4}
\end{eqnarray}
The solution $\Phi=(\Phi_1,\Phi_2)^T$ of this problem is constructed in the same way as for the initial
counterpart. The formulas (\ref{p1}), (\ref{p2}) are combined with the solution (\ref{phi1}) to give
\begin{eqnarray}
\Phi_1 &=& \phi \label{phi1fin} \\
\Phi_2 &=& \frac{1}{m+\sqrt{m^2+\omega~\left(2~n+l+\frac{3}{2}\right)}}
~\left(q_1~\Phi_1-\Phi_1' \right). \label{phi2fin}
\end{eqnarray}
As before, we do not display the explicit form of the latter solution, because it involves very long expressions.
Instead, we show a plot in figure \ref{prob3transf}.
The discrete spectrum admitted by the transformed problem (\ref{bvpd3}), (\ref{bvpd4}) is the same as in
(\ref{isoene}), due to the choice of $\lambda$ in (\ref{lambda}). Before we conclude this example, let us
comment once more on the latter energy $\lambda$. For the numerical value that we chose in our
example, the initial and transformed functions $q_0$ and $q_1$ look very similar, as can be seen by
inspection of figure \ref{q0q1}. This changes once we modify the value of $\lambda$, approaching the
ground state $\epsilon_0$ by employing values of $n$ that are negative and close to zero. We did not
use such values in our calculations, because the resulting expressions would become lengthy due to
very small numbers. Figure \ref{q1ene} shows three examples of functions $q_1$ for different values of
$\lambda$.
Let us now have a look at the effect that the deformation of the transformed function
$q_1$ has on the solutions of the transformed boundary-value problem (\ref{bvpd3}), (\ref{bvpd4}).
For the parameter values used in figure \ref{q1ene} we display three corresponding normalized probability
densities, see figure \ref{phin1ene}.
We observe that the closer $\lambda$ is to the ground state energy $\epsilon_0$, the more
apparent becomes the deformation of the probability density. Before we conclude this paragraph, let 
us briefly comment on higher-order Darboux transformations for the present example, which 
are obtained when instead of (\ref{firsto}) we employ (\ref{phidarb}) for $N>1$. Since in this case 
the expressions for transformed potentials and solutions are in general very complicated, we do not 
show them here, but just indicate how they were obtained. Starting with the case 
$N=2$, we need two 
auxiliary solutions $u_1$ and $u_2$ of (\ref{ssed}) at different energies $\lambda_1, \lambda_2 
\neq \epsilon$ in order to set up the Darboux transformation. The latter solutions are taken from 
(\ref{solosc}), such that instead of (\ref{firsto}) we have
\begin{eqnarray}
\phi &=& \frac{W_{u_1,u_2,\psi_n}}{W_{u_1,u_2}}. \label{2nd}
\end{eqnarray}
This expression, evaluated at $\epsilon=0$ as in (\ref{phienull}), then determines the transformed pseudoscalar 
potential via its parametrizing function $q_1$ in (\ref{q1final}). Figure \ref{q1_2nd} shows an example of 
such a function for particular settings of the parameters, compared to the counterpart (\ref{q0rad}).
The solutions of the transformed Dirac equation are constructed as for the first-order case, that is, by means 
of the formulas (\ref{phi1fin}), (\ref{phi2fin}), where $\phi$ is given in (\ref{2nd}). Higher-order transformations 
can be performed in a completely similar fashion by simply employing a higher value for $N$ in (\ref{phidarb}), 
selecting the auxiliary solutions and their corresponding pairwise different energies, and afterwards following 
the same procedure that was outlined for the second-order case.

\paragraph{Scalar case and position-dependent masses.} The pseudoscalar Dirac potential
in its general form (\ref{v0}) contains the scalar case, as can be seen by comparison with
(\ref{vsca}). In order to obtain a scalar potential from a pseudoscalar counterpart, we must first set
$m=0$ in (\ref{v0}) and redefine $q_0=m+S_0$. As a consequence, we obtain the following
two special cases:
\begin{itemize}
\item[(a)] We can generate a massless Dirac equation for a scalar potential if we further put $m=0$ in (\ref{vsca}).
\item[(b)] A position-dependent mass can be incorporated by allowing the constant $m$ in the scalar
potential (\ref{vsca}) to be spatially dependent, that is, $m=m(x)$.
\end{itemize}
These two special cases can be constructed for the particular example discussed in the previous
paragraph. However, since the results for the transformed potential and the associated solutions would
be very similar, we skip the illustrations for the above mentioned cases in this section.

\section{Discussion}
As shown in the previous sections, the process of constructing Darboux partners of Dirac equations 
for pseudoscalar or scalar potentials that can be expressed through EOPs, is in principle straighforward. 
At this point we would like to discuss certain technical and conceptual difficulties that may arise, and that 
are not easily visible from our example in section 4 or from the preceding general considerations. 
The first issue to be addressed is related to the regularity of the parametrizing functions $q_0$ and 
$q_1$ that generate the pseudoscalar Dirac potential (or the corresponding scalar case). Starting out with 
$q_0$ in its particular form (\ref{q0p}), we observe that the latter form does not have singularities if 
the solution $\hat{q}$ of the linearized equation (\ref{v0q0lin}) does not vanish. If instead of (\ref{q0p}) we 
consider the general solution (\ref{q0final}) for $q_0$, we can avoid singularities by 
additionally imposing that $\int^x 1/\hat{q}^2 ~dt$ is 
bounded from above or below. This choice implies that there is a value for the constant $c$ in 
(\ref{q0final}), such that the denominator in the second term does not vanish. In case the general solution 
of the linearized equation (\ref{v0q0lin}) is available, the solution formula (\ref{q0final}) can be rewritten in 
a more transparent way
\begin{eqnarray}
q_0 &=&  \left[ \frac{d}{dx}~ \log\left(\hat{q}_0 \right) \right] + \frac{1}{
c~\hat{q}_0^2 + \hat{q}_0~\bar{q}_0}, \label{qbar}
\end{eqnarray}
where $\hat{q}_0$ and $\bar{q}_0$ form a fundamental system of (\ref{v0q0lin}). It is clear that the 
same argumentation, as well as a formula analogous to (\ref{qbar}), can be applied to the 
transformed parametrizing function $q_1$, as it obeys equations (\ref{q1p}) and (\ref{q1final}). 
Let us now make one more comment on formula (\ref{qbar}). Since the functions contained in the latter 
formula are solutions of a linear, second-order equation, they are usually special functions. As a 
consequence, the full expression (\ref{qbar}) in general becomes very complicated, such that it 
cannot be used in practical applications. This is especially true when using rationally extended 
potentials that are expressed through EOPs, because these potentials generally take a very long and 
involved form. As a possible consequence, it is common that the integral in (\ref{q1final}) cannot be 
evaluated. This is true for our example presented in section 4, where we 
had to restrict ourselves to the particular form (\ref{q0p}) and (\ref{q1p}) of the functions $q_0$ and 
$q_1$, respectively. The same issue arises when processing Darboux transformations of orders 
higher than one, as mentioned in section 4, where a comparably simple example was considered. 
Let us finally comment on the spectral properties between the initial Dirac problem and its 
Darboux partner. It is well-known that the Darboux transformation of order $N$, when applied to a 
Schr\"odinger problem that admits a discrete spectrum, can generate or remove at most $N$ spectral values 
in the transformed problem, depending on the choice of the auxiliary solutions \cite{djf}. This 
property transfers to the Dirac systems, because both the initial and the transformed problem are 
related to an associated Schr\"odinger problem. As a consequence, the discrete spectra of the Dirac systems 
are modified accordingly.

\section{Concluding remarks}
We have presented a method to generate spectral problems for the Dirac equation involving 
pseudoscalar or scalar potentials that are expressed through EOPs. As can be seen from the 
example we showed in section 4, a principal restriction of our approach lies the feasibility of the 
calculations, which can become very cumbersome due to the length of the expressions involved. 
Since at this point our method is restricted to potentials of pseudoscalar or equivalent type, 
our future research will be dedicated to overcoming the latter restriction, such as to introduce 
potentials related to EOPs in more general potentials and possibly in higher dimensions.

\end{sloppypar}
\end{document}